\documentclass{article}
\usepackage{amsmath}%
\usepackage{amsfonts}%
\usepackage{amssymb}%
\usepackage{graphicx}
\usepackage{braket}
\usepackage{cite}

\newtheorem{theorem}{Theorem}

\newtheorem{lemma}[theorem]{Lemma}

\numberwithin{equation}{section}

\title { Solvability of the generalized  system of stochastic differential equations in driven cavity single mode}
\author{Smail Bougouffa\thanks{
 IMSIU University,
Faculty of  Science, Department of Physics, P.O. Box 90950,
Riyadh 11623, Saudi Arabia, E-mail: sbougouffa@hotmail.com } and Lazhar Bougoffa
\thanks{IMSIU University,
Faculty of  Science, Department of Mathematics, P.O. Box 90950,
Riyadh 11623, Saudi Arabia. E-mail address: lbbougoffa@imamu.edu.sa}
}
\begin{document}

\maketitle
\begin{abstract}\noindent
 A useful approach is investigated in order to analyze a class of a stochastic differential equations that can be encountered in quantum optics problems, especially, in the case of two photon losses on the driven cavity mode. The passage to the ordinary coupled differential equations is presented and the treatment of the obtained coupled system is explored.  Generalization of the problem to stimulate variable coefficients is discussed and the exact solutions are achieved in explicit forms under suitable conditions on the coefficients.
\end{abstract}
\bigskip

\section{Introduction}\label{sec1}
It is well known in quantum optics problems, there are quantum fluctuations associated with the states corresponding to classically well-defined electromagnetic fields. The general explanation of fluctuation phenomena needs the density operator. Nevertheless, it is possible to give an other option but equivalent description in terms of distribution functions~\cite{SZ97,MS07,ZM14}.
It has shown that the extended treatment of quantum statistical phenomena by developing the theory of quasi classical distributions is very interesting field of investigation~\cite{SZ97}. This is of interest for several motivations.
In the begining, the expansion of the quantum theory of radiation to involve nonquantum stochastic effects such as thermal fluctuations is required. This is an important factor in the theory of partial coherence. in addition, the edge between classical and quantum physics is explained by the use of such distribution. The arch type example being the Wigner distribution~\cite{C95}.\\
In addition, the investigation of the border between quantum and classical physics is an attractive issue. Nowhere is this better demonstrated than in quantum optics, where we are regularly encountered with the problem of describing fields which are nearly classical but have significant quantum characters. The coherent states are well appropriated to such studies~\cite{SZ97,MS07,ZM14}.\\
On the other hand, in many quantum optics problems, it is appropriate to illustrate the state of the field in terms of coherent states, rather than with photon number states. This presents some revelations and complications ~\cite{SZ97,MS07,ZM14}, the coherent states are not orthogonal and are over complete. In addition, as we shall notice this over completeness permits us to get a helpful diagonal expansion of the density operator in terms of complex matrix elements $P(\alpha)$. This representation can be understood as a quasi-probability distribution function, whose dynamics can under suitable conditions be described in the form of a Fokker-Planck equation (FPE)\cite{G63, S63,R84,W87,L93, W94}.\\
In addition, from a physical viewpoint, as Fokker Planck equation describes the time evolution of the probability density function of the velocity of a particle under the influence of drag forces and random forces, as in Brownian motion. The equation can be generalized to other observables as well; i.e.,the discrete nonlinear dynamic systems subjected to white random excitation \cite{T63},
the autocorrelation function in semiconductor micro cavities \cite{E1,E2}, gain-swept superradiance (GSS) in an ensemble of two-level atoms \cite{E3} and in cooling and entanglement in cavity optomechanics \cite{E4}, etc....\\
Now in quantum optics, almost problems are governed by the master equation, which is an operator type equation that is not easy to solve. Then we have recourse to use some techniques in order to solve these problems. One of them is the use of the probability distribution that can be appeared from density operator. The reason of the representation of the master equation was to attain c-number differential equations that are equivalent to the operator equations, but are more willingly soluble. In particular, the P-representation is used to estimate the normally ordered correlation functions of the field operators. Furthermore, the P-representation forms a correspondence between the quantum and the classical coherent theory \cite{HJC2002, GZ2004 }. Furthermore, the P-representation is used to transform the master equations into a c-number differential equations called the Fokker-Planck equations \cite{ZM14}, which can be viewed as a stochastic equations. These equations can be solved analytically by direct integration for a specific initial conditions \cite{HJC2002, GZ2004}\\
On the other hand, we will explore the technique to solve the Fokker-Planck equation when direct solutions are not possible. This technique involves stochastic differential equations approach and will be illustrated with an interesting problem in quantum optics, which is concerned with the effect of two-photon losses on the driven cavity mode. We limit our formalism to the case of white noise, i. e., to discrete signal whose samples are regarded as a sequence of serially uncorrelated random variables with zero mean and finite variance. Thus, the obtained system is nonlinear coupled differential equations, which, in general cannot always be analytically solved. We present some treatments of the obtained equations under some suitable conditions. It may, however, be worthwhile if the
physical models can be constructed in such manner that the coupled nonlinear system can either be solved analytically or transformed into another system in which the equations are ordinary and can be decoupled, then solved separately.\\
In this work we will examine a useful approach to solve these obtained coupled nonlinear systems and then discuss
the possibility to enlarge this approach for the case of variable coefficients.

\section{Ito's Stochastic differential equation}\label{sec2}
In the case of initial coherent state ~\cite{T10,B10, BH11, B11,df04,B12}, it has pointed out that the multidimensional Fokker-Planck equation can be analytically solved in some special cases \cite{HJC2002, GZ2004}. In general, the Fokker-Planck equations are not linear and do not admit direct solutions, thus we have recourse to employ other sophistical techniques. One of these techniques is the use of the stochastic differential equations (SDE) approach. This procedure is founded on the fact that for a Fokker-Planck equation with positive diffusion matrix, a set of equivalent stochastic differential equations can be existed. The positive defined diffusion matrix $D_{ij}(\textbf{x}(t),t)$ can always be factorized into the form
\begin{equation}\label{25}
   D_{ij}(\textbf{x}(t),t)=\sum_{k}g_{ik}(\textbf{x}(t),t)g_{kj}^{\dagger}(\textbf{x}(t),t).
\end{equation}
In general, the It\={o} stochastic differential equation (SDE)~\cite{LL07} can be formulated as
\begin{equation}\label{26}
    dx_i=h_i(\textbf{x}(t),t)dt+\sum_{j}g_{ij}(\textbf{x}(t),t)\xi_{j}(t)dt
\end{equation}
where $\textbf{x}=\{x_i, 1\leq i\leq n\}$ is the set of unknowns, the $h_i$ and $g_{ij}$ are arbitrary functions and $\xi_i(t)$ are real independent Gaussian white noise terms with zero mean value $\overline{\xi_i(t)}=0$ and delta-$\delta$ correlated in time
\begin{equation}\label{27}
    \overline{\xi_i(t)\xi_j(t')}=\delta_{ij}\delta(t-t'),
\end{equation}
consequently to the It\={o} 's lemma \cite{LL07}, the set of stochastic differential equations, which is associated to the general form of Fokker-Planck equation ~\cite{DW80, AM14}, can be read as
\begin{equation}\label{28}
dy=\sum_{i}\frac{\partial y}{\partial x_i}\Big(h_idt+\sum_jg_{ij}\xi_{j}(t)dt\Big)+\frac{1}{2}\sum_{k,l,m}\frac{\partial^2y}{\partial x_k\partial x_l}g_{km}g_{lm}dt+\frac{\partial y}{\partial t}dt
\end{equation}
where $y=y(x_k,t)$ is a smooth function of the unknown variables $x_k$.\\
 On the other hand, the stochastic differential equations can be treated by numerical simulation techniques or by analytic methods when they are linear ~\cite{DW80, AM14}. But if they are non linear equations, the situation will become more complicated and needs to investigate new approaches. In the following, we will limit our self to an interesting problem that can be encounter when we deal with the effect of two-photon losses on the driven cavity mode in quantum optics.

\subsection{Two-photon losses effect on the driven cavity
mode}\label{sec 2.1} Assume that the cavity mode, considered in the
previous example, is in addition damped by two-photon
losses~\cite{DW80, AM14}, e.g., due to a two-photon absorption.
Then, its corresponding interaction Hamiltonian can be read in the
interaction picture as
\begin{eqnarray}\label{38}
  H_{3} &=& \hbar\sum_{\bf{k}}\left( G_{\bf{k}} b_{\bf{k}}^{\dag}a^2
  + G_{\bf{k}}^{*}(a^{\dag})^{2}b_{\bf{k}}\right).
\end{eqnarray}
Assume the heat bath is at zero temperature, the master equation of the system can be written as ~\cite{T10,B10, BH11, B11,df04,B12}
\begin{eqnarray}\label{39}
    \frac{\partial}{\partial t}\rho &=&\Big[-i\delta a^\dag a+Ea^\dag-E^*a,\rho\Big]
    -\frac{1}{2}\kappa \Big([a^{\dag 2},a^{2}\rho]+[\rho a^{\dag 2},a^{2}]\Big)\nonumber\\&-&\frac{\gamma}{2}\Big([a^{\dag},a\rho(t)]+
  [\rho(t)a^{\dag},a]\Big)
\end{eqnarray}
where $\kappa$ is the photon-photon interaction term(the two-photon loss coefficient). The obtained density matrix can be converted to Fokker-Planck equation using the previous treatment. However, the diffusion matrix of FPE is not always positive defined. Furthermore, as the stochastic processes are independent, the corresponding stochastic differential equations for $\alpha$ and $\alpha^*$ are not complex conjugate. To remedy this problem we will use the positive P representation. In this situation, the stochastic differential equations can be obtained only by using the positive P-representation; the usual P-representation does not work \cite{ZM14}. The positive P-representation is defined as follows
\begin{equation}\label{40}
    \rho=\int_{\mathcal{D}}\Lambda(\alpha,\beta)P(\alpha,\beta)d^2\alpha d^2\beta,
\end{equation}
where
\begin{equation}\label{41}
    \Lambda(\alpha,\beta)=\frac{\ket{\alpha}\bra{\beta^*}}{\langle\beta^*|\alpha\rangle},
\end{equation}
and $(\alpha,\beta)$ vary independently over the whole complex plane $\mathcal{D}$. The projection operator $\Lambda(\alpha,\beta)$ is analytic in $(\alpha,\beta)$.

Subsequent the standard method, using the positive $P$ representation, we convert the master equation into a FPE, then, it can be written in the interaction picture as
\begin{eqnarray}\label{39}
    \frac{\partial}{\partial \alpha}P(\alpha,\beta,t)=&\Bigg(&\frac{\partial}{\partial \alpha}\Big((\frac{1}{2}\gamma+i\delta)\alpha + \kappa \alpha^2\beta^*-E\Big)\nonumber \\&+&\frac{\partial}{\partial \beta^*}\Big((\frac{1}{2}\gamma-i\delta)\alpha^*+\kappa \beta^{*2}\alpha-E^*\Big)\nonumber\\
    &+&\frac{1}{2}\frac{\partial^2}{\partial \alpha^2}(-\kappa \alpha^2)+\frac{1}{2}\frac{\partial^2}{\partial \beta^{*2}}(-\kappa \beta^{*2})\Bigg)P(\alpha,\beta, t).
\end{eqnarray}
The associated It\={o}'s stachastic differential equations can be obtained
\begin{eqnarray}
  \frac{d\alpha}{dt} &=& a\alpha -\kappa \alpha^2\beta^*+E+g_{11}\xi_{1}(t) \label{40}\\
  \frac{d\beta^*}{dt} &=& a^*\beta^*-\kappa \beta^{*2}\alpha+E^*+g_{22}\xi_{2}(t)\label{41}
\end{eqnarray}
where $a=-(\frac{1}{2}\gamma +i\delta)$ and $\xi_i(t)$ are independent Gaussian noise terms with zero means that satisfying the nonzero correlations Eq.(\ref{27}). The coefficients $g_{11}$ and $g_{22}$ are the diagonal matrix elements of the matrix $\textbf{g}$ that can be deduced from the diffusion matrix $\textbf{D}=\textbf{g}\textbf{g}^{T}=diag[-\kappa \alpha^2, -\kappa \beta^{*2}]$. Then, $g_{11}=i\sqrt{\kappa}\alpha$ and $g_{22}=-i\sqrt{\kappa}\beta^*$.\\
The equations (\ref{40}, \ref{41}) constitute a coupled no-linear differential equations of first order. In general, this system is solved numerically or transformed to high order separable differential equations, which may also be treated numerically. In order to avoid the numerical volume, we will investigate an analytic treatment that can be useful to solve this nonlinear coupled system.

\subsection{Stationary deterministic solutions}\label{sec2.2}
In order to explore the stability of the solutions of the coupled system, the deterministic stationary solutions of the system (\ref{40}, \ref{41}) is required and can be obtained by neglecting the noises terms and assuming the time derivatives equal zero. Then
\begin{eqnarray}\label{42}
  \alpha^{s} &=& -\frac{a^*Z_0+E^*}{\kappa Z_0^2} \\
  \beta^{*s} &=& Z_0
\end{eqnarray}
where $Z_0$ is a root of the third order equation
\begin{equation}\label{43}
    \kappa EZ_0^3+a^*(a-a^*)Z_0^2+E^*(a-2a^*)Z_0-E^{*2}=0.
\end{equation}
In general, these solutions are not stable except for some special cases. In particular, if the classical coherent field is ignored $E=0$, then the stationary solution is $\alpha=\beta=0$ and this solution is stable. Further, in the case of $\gamma=0$ and $\delta=0$; i.e., the cavity mode is in resonance with the classical coherent filed and the vacuum fluctuations are negligible, then the stationary solution is $\alpha^s=\beta^{*s}=(\frac{E}{\kappa})^{\frac{1}{3}}$. For $E>0$, this solution is stable. On the other hand, the solutions $\alpha^s=-\beta^{*s}=i^{\frac{2}{3}}(\frac{E}{\kappa})^{\frac{1}{3}}$ are unstable solutions for $E>0$. However, it is this sort of rearrangement of the deterministic nonlinear dynamics in the extended phase space which can direct to irregular solutions (unstable trajectories) in numerical treatments of the stochastic differential equations when the quantum noises are large.

\subsection{Linear noise approximation}\label{sec2.3}
If the master equation possesses nonlinear transition rates, it may be impractical to solve it systematically and we recourse to other technical approximations. One of them is the linear noise approximation, which plays important role in many different physics models \cite{E5, AS}.
In the case of small noise terms, the linear noise approximation can
be used, in which the fluctuations are linearized around the steady
state solution. On the other hand, we should be conscious that the
obtained results will be appropriate just in the framework of this
approximation. However, except the problem of passage from one
stationary state to another in bistable systems, this is mainly what
is obvious well.\\ Let's consider now the previous coupled system
(\ref{40}, \ref{41}). It is clear that this equation does not
include any very apparent small
 noise factor. Nevertheless, a large driving field regime can be attained by assuming\\
\begin{equation}\label{44}
    \kappa=\frac{b}{E^2}, \quad \alpha=xE, \quad \beta^*=yE, \quad E^*=E
\end{equation}
then
\begin{eqnarray}
  \frac{dx}{dt} &=& ax -b x^2y+1+i\frac{\sqrt{b}}{E}x\xi_{1}(t), \label{45}\\
  \frac{dy}{dt} &=& a^*y-b y^{2}x+1-i\frac{\sqrt{b}}{E}y\xi_{2}(t),\label{46}
\end{eqnarray}
This clearly shows that in the regime of large driving field, the small linearization approximation can be applied.
\section{Solvability of  the  coupled ordinary equations}\label{sec3}
In order to investigate the deterministic solution of the stochastic equations, it is interesting to explore some real physical problem where the stochastic differential equation can be converted to an ordinary coupled differential equations. In the framework of large driving field, the noise terms can be ignored.
On the other hand, one of the main problem of mathematics~\cite{BB14} appears when $a, \ b$  are analytic functions and are added to the original system. Now the new problem, incorporating the above assumptions, is generated by a coupled differential equations,
\begin{eqnarray} \left\{
\begin{array}{rcl}
\frac{dx}{dt}&=&a(t)x+b(t)x^{2}y+f(t),\\\\
\frac{dy}{dt}&=&c(t)y+d(t)y^{2}x+g(t).
\end{array}
\right.\label{43}
\end{eqnarray}
A question which arises naturally is under what conditions on the
functions $a(t), \ b(t), \ c(t)$ and $d(t)$ does the given system
have an explicit solution?\\ In this section, we will present a
direct approach to solve the general coupled model by considering
the following two important cases.
\subsection{The homogeneous
coupled system}\label{sec3.1} First of all, we begin our approach by
considering the following homogeneous coupled system
\begin{eqnarray} \left\{
\begin{array}{rcl}
\frac{dx}{dt}&=&a(t)x+b(t)x^{2}y,\\\\
\frac{dy}{dt}&=&c(t)y+d(t)y^{2}x.
\end{array}
\right.\label{44}
\end{eqnarray}
Multiplying both sides of the first and second equation of system
(\ref{44}) by $y$ and $x,$  respectively, we get
\begin{equation}
\label{45}y\frac{dx}{dt}=a(t)xy+b(t)x^{2}y^{2}
\end{equation}
and
\begin{equation}
\label{46}x\frac{dy}{dt}=c(t)xy+d(t)x^{2}y^{2}.
\end{equation}
Adding Eq. (\ref{45}) and Eq. (\ref{46}) together, we obtain
\begin{equation}
\label{47}y\frac{dx}{dt}+x\frac{dy}{dt}=\left(a(t)+c(t)\right)xy+
\left(b(t)+d(t)\right)x^{2}y^{2}.
\end{equation}
Now let
\begin{equation}
\label{48}z=xy.
\end{equation}
Hence, Eq. (\ref{47}) becomes
\begin{equation}
\label{49}\frac{dz}{dt}=A(t)z+B(t)z^{2},
\end{equation}
where $A(t)=a(t)+c(t)$ and $B(t)=b(t)+d(t),$ which is a Bernoulli
differential equation. The substitution that is needed to solve this
Bernoulli equation is
\begin{equation}
z=\frac{1}{u}.
\end{equation}
A set of solutions to Eq. (\ref{49}) is then given by
\begin{equation}
\label{50} z=\frac{e^{\int A(t)dt}}{-\int B(t)e^{\int
A(t)dt}dt+\alpha},
\end{equation}
where $\alpha$ is a constant of integration.\\
Inserting Eq. (\ref{48}) into system (\ref{44}) to get
\begin{equation}
\frac{dx}{dt}=\left(a(t)+b(t)z(t)\right)x
\end{equation}
and
\begin{equation}
\frac{dy}{dt}=\left(c(t)+d(t)z(t)\right)y.
\end{equation}
Consequently, 
\begin{equation}
\label{51}x(t)=\beta e^{\int\left(a(t)+b(t)z(t)\right)dt}
\end{equation}
and
\begin{equation}
\label{52}y(t)=\gamma e^{\int\left(c(t)+d(t)z(t)\right)dt},
\end{equation}
where $\beta$ and $\gamma$ are two constants of integration, and  the function $z$ is given by Eq. (\ref{50}).\\
Thus we have proved the following result on the separation of this
system.
\begin{lemma}
The two equations of system (\ref{43}) are decoupled and solvable
separately, and the solutions are given by Eq. (\ref{51}) and Eq.
(\ref{52}).
\end{lemma}

\subsection{The non-homogeneous coupled system}\label{sec3.2} In
what follows, we will examine the nonlinear system  (\ref{43}).
\subsubsection{$x(t)$ and $y(t)$ are proportional } The system
(\ref{43}) can be transformed into another system in which the
equations are decoupled and transformed into one equation. This
separability can be obtained if
\begin{equation}
\label{53}y(t)=\varphi(t)x(t),
\end{equation}
where $\varphi(t)$ is unknown function.\\
The substitution of (\ref{53}) into system (\ref{43}) gives a system
which we write as:
\begin{eqnarray} \left\{
\begin{array}{rcl}
\frac{dx}{dt}&=&a(t)x+b(t)\varphi(t)x^{3}+f(t),\\\\
\frac{dx}{dt}&=&\frac{c(t)\varphi(t)-\varphi'(t)}{\varphi(t)}x+d(t)\varphi(t)x^{3}+\frac{g(t)}{\varphi(t)}.
\end{array}
\right.\label{54}
\end{eqnarray}
Equating coefficients of like terms of system (\ref{54}), we get
\begin{equation}
\label{55}\frac{c(t)\varphi(t)-\varphi'(t)}{\varphi(t)}=a(t),
\end{equation}
\begin{equation}
\label{56}b(t)=d(t)
\end{equation}
and
\begin{equation}
\label{57}\varphi(t)=\frac{g(t)}{f(t)}.
\end{equation}
Eq. (\ref{55}) gives
\begin{equation}
\label{58}\varphi(t)=\lambda e^{\int(c(t)-a(t))dt},
\end{equation}
where $\lambda$ is a constant.\\
Thus we have proved the following result on the separation of this
system.
\begin{lemma}
The system (\ref{43})  can be reduced to the Abel equation of the
first kind~\cite{A03}
\begin{equation}
\label{59}\frac{dx}{dt}=f_{0}(t)+f_{1}(t)x+f_{2}(t)x^{3},
\end{equation}
where $f_{0}=f, \ f_{1}=a$ and $f_{2}=b\varphi,$ such that
\begin{equation}
\label{60}y=\frac{g(t)}{f(t)}x
\end{equation}
if and only if the following conditions
\begin{equation}
 \label{61}\frac{g(t)}{f(t)}=\lambda
e^{\int(c(t)-a(t))dt}
\end{equation}
and Eq.(\ref{56}) are satisfied.
\end{lemma}
For the previous problem, we have $b=d$, $a=c^*=-(\frac{\gamma}{2}+i\delta)$ and $f=g=1$, then $y=x$ and the reduced equation takes the form
\begin{equation}\label{62}
  \frac{dx}{dt}=1+ax+bx^{3},
\end{equation}
where $\lambda=e^{-\int 2i\delta dt}.$\\

Listed below are some special cases when the Abel equation Eq.
(\ref{59}) is solvable.
\begin{enumerate}
  \item If the functions $f_{n}(n=0,1,2)$ are
proportional that is $f_{n}(t)= k_{n}h(t),$ then Eq. (\ref{59}) is a
separable equation. Therefore,
\begin{equation}
 \int\frac{dx}{k_{0}+k_{1}x+k_{2}x^{3}}=\int h(t)dt.
\end{equation}
  \item If $f_{0}=k_{0} t^{-n-2}, \
f_{1}(t)=\frac{k_{1}}{t}$ and $f_{2}(t)=k_{2}t^{2n+1}.$ The
substitution $z(t)=t^{n+1} x(t)$ leads to a separable equation
\begin{equation}
 t \frac{dz}{dt}=k_{0}+(k_{1}+n+1)z+k_{2}z^{3}.
\end{equation}
  \item If $f_{0}=k_{0} t^{2m}, \ f_{1}(t)=k_{1}
t^{m+n}$ and $f_{2}(t)=k_{2}t^{3n-m},$ then the  Abel equation Eq.
(\ref{59}) can be reduced with the substitution $x(t)=t^{m-n} z(t)$
to a separable equation
\begin{equation}
 t^{-n-m} \frac{dz}{dt}=k_{0}+k_{2}z^{3}.
\end{equation}
  \item If the Abel equation Eq. (\ref{59}) is
generalized homogeneous, that is $f_{0}=k_{0} t^{-n-2}, \ f_{1}(t)=
\frac{k_{1}}{t}$ and $f_{2}(t)=k_{2}t^{2n+1},$ then it can be
reduced with the substitution $z(t)=t^{n+1} x(t)$ to a separable
equation
\begin{equation}
 t \frac{dz}{dt}=k_{0}+(k_{1}+n+1)z+k_{2}z^{3}.
\end{equation}
\end{enumerate}

\subsubsection{General case}

Let's start off by introducing the following transformations
\begin{equation}
\label{62}x(t)=\psi(t)x_{h}(t)\  \ \mbox{and} \ y(t)=\chi
(t)y_{h}(t),
\end{equation}
where $\psi(t)$  and $\chi$ are two  differentiable functions to be
determined and $(x_{h}(t), y_{h}(t))$ is a solution of its
corresponding homogeneous system (\ref{44}).\\Substituting
(\ref{62}) into (\ref{43}), we get
\begin{eqnarray} \left\{
\begin{array}{rcl}
x_{h}\psi'(t)+\psi(t)\dfrac{dx_{h}}{dt}&=&a(t)\psi(t)x_{h}+b(t)\psi^{2}(t)\chi(t) x^{2}_{h}y_{h}+f(t),\\\\
y_{h}\chi'(t)
+\chi(t)\dfrac{dy_{h}}{dt}&=&c(t)\chi(t)y_{h}+d(t)\psi(t)\chi^{2}(t)x_{h}y^{2}_{h}+g(t).
\end{array}
\right.\label{64}
\end{eqnarray}
If we choose $\chi(t)=\frac{1}{\psi(t)},$ and taking into account
that  $(x_{h}(t), y_{h}(t))$ is a solution of the homogeneous system
(\ref{44}), then (\ref{64}) reduces to
\begin{eqnarray} \left\{
\begin{array}{rcl}
x_{h}\psi'(t)&=&f(t),\\\\
y_{h}\left(\dfrac{1}{\psi(t)}\right)'&=&g(t),
\end{array}
\right.\label{65}
\end{eqnarray}
or
\begin{eqnarray} \left\{
\begin{array}{rcl}
\psi'(t)&=&\dfrac{f(t)}{x_{h}(t)},\\\\
\left(\dfrac{1}{\psi(t)}\right)'&=&\dfrac{g(t)}{y_{h}(t)},
\end{array}
\right.
\end{eqnarray}
that is
\begin{eqnarray} \left\{
\begin{array}{rcl}
\psi'(t)&=&\dfrac{f(t)}{x_{h}(t)},\\\\
-\dfrac{\psi'(t)}{\psi^{2}(t)}&=&\dfrac{g(t)}{y_{h}(t)}.
\end{array}
\right.\label{66}
\end{eqnarray}
Solving system (\ref{66}) for $\psi(t),$ we get $\psi(t)=
\sqrt{-\frac{f(t)}{g(t)}\frac{y_{h}(t)}{x_{h}(t)}}, \ \mbox{where}
\ \frac{f(t)}{g(t)}\frac{y_{h}(t)}{x_{h}(t)}<0.$\\
Thus we have
\begin{lemma}
The system (\ref{43}) has a solution of the form
\begin{equation}
\label{67}(x(t), \ y(t))=(\psi(t)x_{h}(t),  \
\frac{1}{\psi(t)}y_{h}(t)),
\end{equation}
where
\begin{equation}
\label{68}\psi(t)=\sqrt{-\frac{f(t)}{g(t)}\frac{y_{h}(t)}{x_{h}(t)}}
\end{equation}
provided $\frac{f(t)}{g(t)}\frac{y_{h}(t)}{x_{h}(t)}<0$ and
$(x_{h},\ y_{h})$ is given by Eqs. (\ref{51})-(\ref{52}).
\end{lemma}
\section{Conclusion}\label{sec4}
We have explored some techniques to analysis the quantum stochastic differential equations, which are generated in the case of driven single cavity mode in different reservoirs. One of the mean results resides essentially in the conversion of the stochastic equations to ordinary differential equations. Generalization coverage of the nonlinear coupled differential equation is presented. These results stimulate a sequence of questions of mathematical as well as physical consequence. Ideally one would like to be capable of predicting the behavior of paths for any set of initial conditions and parameter values. This is very much an interesting question in general. The quantum optics models offer an important source in the nonlinear aspects, and have motivated the development of techniques to examine more and more difficult and higher dimensional models.\\

\end{document}